\documentclass[journal,comsoc]{IEEEtran}
\usepackage[T1]{fontenc}
\usepackage{cite}
\usepackage[pdftex]{graphicx}
\usepackage{multirow,nicematrix}
\usepackage{amsmath}
\usepackage{bm}
\usepackage{xcolor}
\usepackage[labelformat=simple]{subcaption}
\usepackage{booktabs}
\usepackage{microtype}
\usepackage{graphics}

\newcommand{\revision}[1]{{#1}}

\usepackage{tikz}
\usepackage{pgfplots}
\usetikzlibrary{calc}
\usetikzlibrary{positioning}
\pgfplotsset{compat=newest}
\usepackage{url}
\usepackage{standalone}
\newcommand\tikzscale{1}
\providecommand{\datapath}{figures}
\definecolor{col1}{rgb}{0.0000,0.4470,0.7410}%
\definecolor{col2}{rgb}{0.8500,0.3250,0.0980}%
\definecolor{col3}{rgb}{0.9290,0.6940,0.1250}%
\definecolor{col4}{rgb}{0.4940,0.1840,0.5560}%
\definecolor{col5}{rgb}{0.4660,0.6740,0.1880}%
\definecolor{col6}{rgb}{0.3010,0.7450,0.9330}%
\tikzset{%
  >=latex%
}
\begin{document}
\title{Wireless Channel Charting: \\ Theory, Practice, and Applications}

\author{Paul~Ferrand,~\IEEEmembership{Senior~Member,~IEEE,}
        Maxime~Guillaud,~\IEEEmembership{Senior~Member,~IEEE,}
        Christoph~Studer,~\IEEEmembership{Senior~Member,~IEEE,}
        Olav~Tirkkonen,~\IEEEmembership{Fellow,~IEEE}
}



\maketitle
\begin{abstract}
Channel charting is a recently proposed framework that applies dimensionality reduction to channel state information (CSI) in wireless systems with the goal of associating a pseudo-position to each mobile user in a low-dimensional space: the \emph{channel chart}. Channel charting summarizes the entire CSI dataset in a self-supervised manner, which opens up a range of applications that are tied to user location. In this article, we introduce the theoretical underpinnings of channel charting and present an overview of recent algorithmic developments and experimental results obtained in the field. We furthermore discuss concrete application examples of channel charting to network- and user-related applications, and we provide a perspective on future developments and challenges as well as the role of channel charting in next-generation wireless networks.
\end{abstract}

\IEEEpeerreviewmaketitle

\section{Introduction}

\IEEEPARstart{M}{odern} wireless communications systems rely on ever-increasing bandwidth and number of antennas to enable better service.
This trend results in a growing number of parameters required to represent the state of the wireless channel, even though channel modeling studies have demonstrated that the wireless channel can be characterized by a comparatively small set of parameters, such as the locations of antennas and environmental scatterers.

Communication protocols and applications increasingly rely on knowledge about the geographical location of the user equipments (UEs) to support high-level network functions, such as handover between base stations and proximity detection~\cite{Kasparick_TVT2016}.
Location-based services are expected to bring a significant part of the value provided by 6G networks~\cite[Chap.~3]{tong_zhu_2021}.
Unfortunately, existing positioning approaches are either imprecise or
require a significant amount of dedicated resources to obtain feedback from the UE~\cite{Alamu2021}.
They are also not universally applicable: UEs may not have global navigation satellite system (GNSS) positioning capabilities, they may be indoors or without satellite coverage, or they may forbid sharing the position with a third party.

For many location-based services, absolute position information is in fact not required: a {\it pseudo-location} that accurately characterizes short-distance neighborhood relationships may be sufficient. Capitalizing on this, channel charting (CC) has recently emerged as an alternative to classical localization techniques that alleviates some of the aforementioned drawbacks~\cite{Studer2018}.
CC consists of learning a mapping between CSI samples and points in a so-called \emph{channel chart}, which is defined as a compact set in a low-dimensional Euclidean space.
This mapping is constructed such that nearness between two points on the channel chart indicates similarity between the corresponding CSI samples, their propagation conditions, and thus their proximity in the physical space.
Desirable properties of the channel chart are consistency in time and across the UEs, and low dimensionality.

\begin{figure*}[tp]
\centering
    \includegraphics[width=1.5\columnwidth]{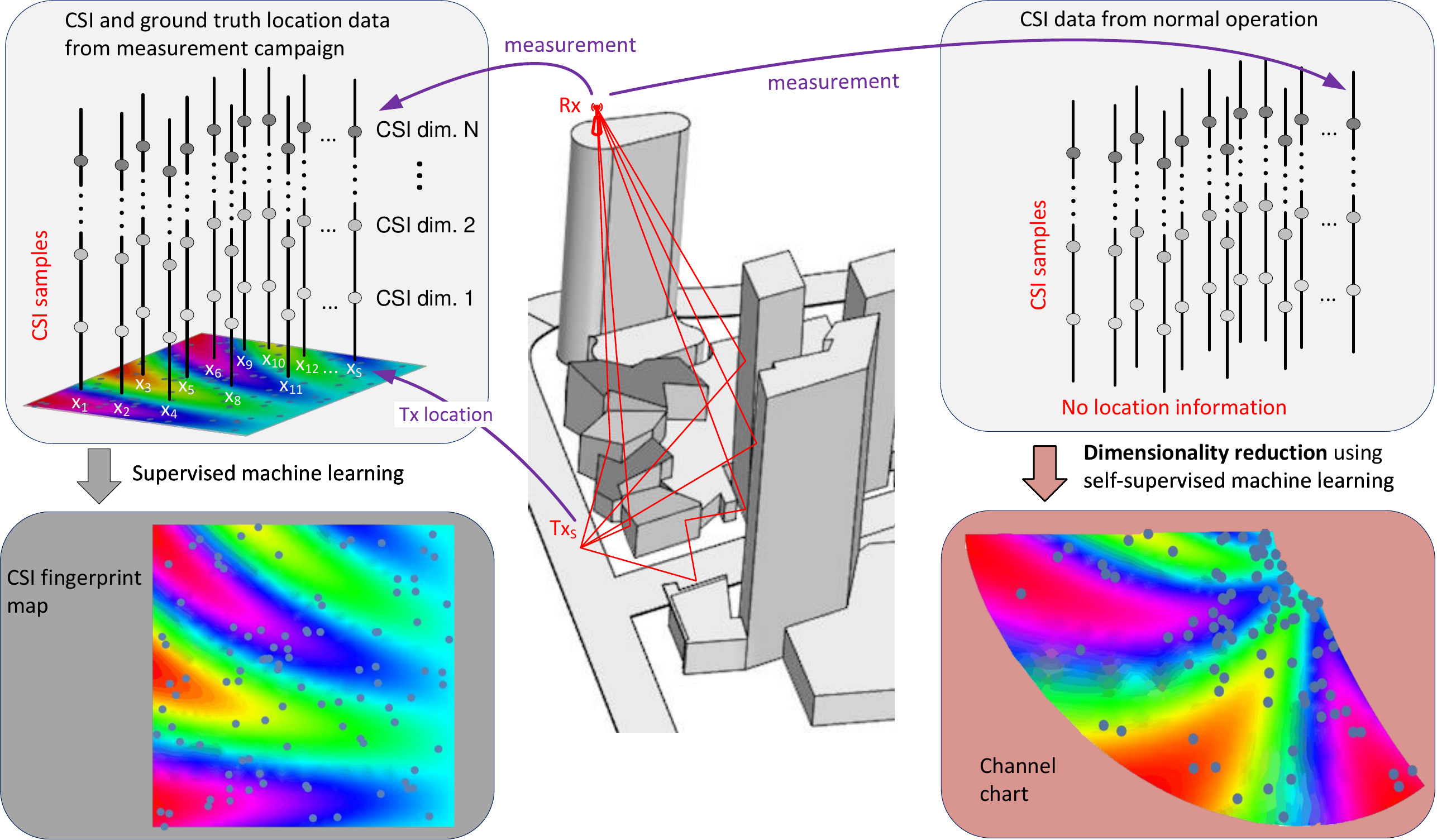}
    \caption{\revision{
    CSI fingerprinting vs.\ channel charting (CC). CSI fingerprinting (left) requires CSI samples labeled with location information 
    for training.
    CC (right) uses unlabeled CSI samples only, thus avoiding costly measurement campaigns; the resulting chart is locally consistent but not necessarily 
    globally accurate.}}
    \label{fig:fingerprinting_vs_CC}
\end{figure*}

Every CSI estimate gives rise to a sample represented by a point in the space of CSI values.
In order to learn a meaningful mapping between a set of samples in the CSI space and their representations \revision{in the channel chart},
CC leverages self-supervised dimensionality reduction techniques~\cite{VanDerMaaten2009}.
Self-supervision is a distinctive characteristic compared to classical positioning techniques, as it allows channel chart generation purely from collected CSI samples---no reference location information from GNSS or from ad-hoc surveys is required.
As a consequence of self-supervision, a channel chart pseudo-position only constitutes relative position information w.r.t. other CSI samples and not an absolute position in physical space.
\revision{This important aspect is illustrated in Fig.~\ref{fig:fingerprinting_vs_CC}, which provides a side-by-side comparison between CSI fingerprinting and channel charting.}
\revision{The lack of global geometric consistency in channel charts} can be alleviated by resorting to semi-supervised CC approaches, \revision{in which a subset of transmit locations are known.}
Being entirely data-driven, CC naturally handles complex propagation geometries while avoiding the complexity associated with possibly mismatched geometric models.

Many radio resource management (RRM) tasks can be designed to utilize a relative position within the channel chart~\cite{Bourdoux2020}.
The pseudo-position enables infrastructure base stations (BS) to perform context-based RRM, such as SNR prediction or pilot sequence distribution~\cite{Kazemi2021,Ribeiro2020}. 
CC can answer questions such as ``are two users close?'' or ``is the user moving towards an area with poor service?'' with minimal protocol overhead, while preserving location privacy and processing only CSI data that is acquired at the BS anyway.

In this article we provide an intuitive 
idea of the principles of CC (Secs.~\ref{sec:manifold} and \ref{sec:dimreduction}), and of the related algorithmic and practical implementation aspects (Sec.~\ref{sec:charting}).
Since the introduction of CC in \cite{Studer2018}, several research groups have demonstrated that meaningful channel charts can be practically obtained using data originating from both simulated~\cite{Studer2018} and measured propagation environments~\cite{Ferrand2021};
these will be discussed in Sec.~\ref{sec:experiments}.
\revision{We showcase two concrete examples that leverage CC and  provide a list of potential applications (Sec.~\ref{sec:applications}), and we discuss challenges, extensions, and future research perspectives (Sec.~\ref{sec:futuredirections}).}
A regularly updated collection of CC-related publications, code, and datasets may be found on the \emph{Channel Charting Resources} website~\cite{channelchartingresources_website}.

\section{The Wireless Channel Manifold and its Sampling}
\label{sec:manifold}

Wireless signals propagate from a transmitting antenna to a receiving antenna according to Maxwell's equations.
The set of parameters governing these equations include the positions of the transmitter (Tx) and receiver (Rx) antenna elements, the carrier frequency, as well as the nature and positions of scattering objects in the environment.
The effective channel response experienced by a user is a continuous function of these parameters.

Consider, e.g.,
a set-up with a static BS receiver 
equipped with an antenna array in a static scattering environment, 
and a single-antenna mobile user transmitter located anywhere in a two-dimensional (2D) coverage area.
The complex-valued vector of channel responses measured at the Rx antennas at a given instant in time (a \emph{CSI sample}) is 
a point in the \emph{CSI space}, continuously varying as a function of the Tx position.
This is the \emph{antenna array manifold} representation of direction-finding algorithms~\cite{Schmidt1986}.
In our example with static scatterers, the locus of possible CSI samples thus forms a continuous image of the 2D coverage area in the CSI space.
Not all CSI samples at the receiver are possible: only those CSI vectors that lie on the image of the coverage area can occur in reality.

This image can be understood as a crumbled, folded, and stretched surface embedded in the high-dimensional CSI space, representing the coverage area.
It is not necessarily a manifold---the image may intersect with itself if different locations yield the same CSI.
However, such nonuniqueness is not common, and we shall see that it helps to think of the image as a manifold.
A CSI sample is a point from this surface.
Channel charting is based on the idea that 
with a sufficiently large number of CSI samples, one may identify the dominant features of the image of the coverage area,
and attempt to unfold it into a compact set 
in a low-dimensional space.

Consider now the situation from a channel modeling perspective.
Tractable models use simplified plane-wave approximations of wave equations, where the channel depends on a relatively small subset of the most essential environmental parameters.
Absorption, reflection, and diffraction, as well as scattering effects within the environment \revision{are typically considered for each wave.}
\revision{
While such models postulate a propagation model and attempt to estimate its parameters, CC \emph{jointly} learns the model and its parameters. 
This key difference enables CC to operate under complex and realistic propagation conditions, affected by multi-path, reflection, diffraction, shadowing, etc.
These effects are contained in the data and used in training the channel chart, rather than considered 
to be corrupting an assumed propagation model.}

Reality is of course more complex than a static scattering scenario.
However, most objects significantly affecting propagation are static (such as buildings) or have limited mobility (such as cars).
In an environment with some mobility, the image of the 2D coverage area would not be a surface, but an object of higher dimension.
The relative importance of added dimensions will depend on the impact of the underlying movement on the array manifold.

CSI dimensionality has tremendously increased in recent years.
Transmission bandwidths are larger, and base stations and UEs have more antenna elements;
contemporary BSs commonly operate with 64 antennas over a 20\,MHz bandwidth with around 1200 subcarriers.
The dimension of each CSI sample containing all subcarriers and antennas would then easily be in tens of thousands.
During normal operation of a radio access network (RAN), the channel from a user may be measured hundreds of times per second.
CC
summarizes it into a lower-dimensional representation of the data akin to the position of a user on a map.
\revision{This information can be used as a pseudo-position at higher layers for multiple purposes, as discussed in Sec.~\ref{sec:applications}.}

\section{Manifold Learning via Dimensionality Reduction}
\label{sec:dimreduction}

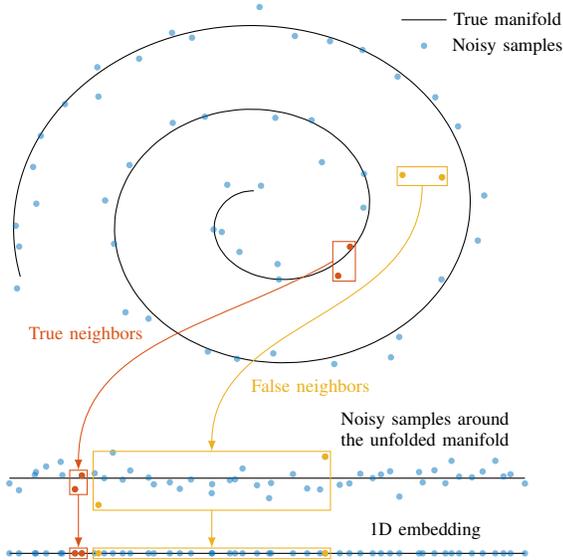
\begin{figure}[tb!]
    \renewcommand\tikzscale{1}
    \centering
    \small

\pgfplotstableread{\datapath/spiral_noise.data}{\spiralnoisetable}
\pgfplotstableread{\datapath/unrolled_spiral_noise.data}{\unrollednoisetable}

\begin{tikzpicture}[transform shape, scale=\tikzscale]
  \scriptsize
  \begin{axis}[
    name=spiral,
    enlargelimits=false,
    ticks=none,
    axis lines=none,
    xmin=-0.8, xmax=0.8,
    ymin=-0.8, ymax=0.8,
    legend style={at={(1.1,1.00)}, anchor=north east, draw=none, fill=none}
  ]
  \addplot+[mark=none, color=black, smooth] table {\datapath/spiral.data};
  \addlegendentry{True manifold};
  \addplot+[only marks, mark=*, mark size=1, opacity=0.5, mark options={fill=col1, color=col1}]  table {\spiralnoisetable};
  \addlegendentry{Noisy samples};
  \coordinate (spiral-1) at (0.22012955461880554, -0.22696123769060297);
  \coordinate (spiral-2) at (0.2572930765488756, -0.11853655671881475);
  \coordinate (spiral-jump-1) at (0.42, 0.15);
  \coordinate (spiral-jump-2) at (0.5424192593011172, 0.1412292442587747);
  \end{axis}
  
  \draw[col2, fill=col2] (spiral-1) circle (1pt);
  \draw[col2, fill=col2] (spiral-2) circle (1pt);
  \draw[col3, fill=col3] (spiral-jump-1) circle (1pt);
  \draw[col3, fill=col3] (spiral-jump-2) circle (1pt);
  \path let \p1 = (spiral-1) in coordinate (spiral-sw) at (\x1 - 2, \y1 - 2);
  \path let \p2 = (spiral-2) in coordinate (spiral-ne) at (\x2 + 2, \y2 + 2);
  \draw[col2] (spiral-sw) rectangle (spiral-ne);
  
  \path let \p1 = (spiral-jump-1) in coordinate (spiral-jump-sw) at (\x1 - 2, \y1 - 4);
  \path let \p2 = (spiral-jump-2) in coordinate (spiral-jump-ne) at (\x2 + 2, \y2 + 4);
  \draw[col3] (spiral-jump-sw) rectangle (spiral-jump-ne);

  \begin{axis}[
    name=unrolled,
    yshift=-3.5cm,
    enlargelimits=false,
    ticks=none,
    axis lines=none,
    ymin=-0.8, ymax=0.8
  ]
  \addplot+[mark=none, color=black, smooth] table {\datapath/unrolled_spiral.data};
  \addplot+[col1, only marks, mark=*, mark size=1, opacity=0.5, mark options={fill=col1, color=col1}]  table[x=0, y=1] {\unrollednoisetable};
  \coordinate (unrolled-1) at (1.8876326887658077, -0.04135223354032603);
  \coordinate (unrolled-2) at (1.9556583067767028, 0.00994523397222574);
  \coordinate (unrolled-jump-1) at (2.12, -0.1);
  \coordinate (unrolled-jump-2) at (4.368817359432314, 0.08041996464637106);
  \end{axis}
  
  \draw[col2, fill=col2] (unrolled-1) circle (1pt);
  \draw[col2, fill=col2] (unrolled-2) circle (1pt);
  \path let \p1 = (unrolled-1) in coordinate (unrolled-sw) at (\x1 - 2, \y1 - 2);
  \path let \p2 = (unrolled-2) in coordinate (unrolled-ne) at (\x2 + 2, \y2 + 2);
  \draw[col2] (unrolled-sw) rectangle (unrolled-ne);
  
  \draw[col3, fill=col3] (unrolled-jump-1) circle (1pt);
  \draw[col3, fill=col3] (unrolled-jump-2) circle (1pt);
  \path let \p1 = (unrolled-jump-1) in coordinate (unrolled-jump-sw) at (\x1 - 2, \y1 - 2);
  \path let \p2 = (unrolled-jump-2) in coordinate (unrolled-jump-ne) at (\x2 + 2, \y2 + 2);
  \draw[col3] (unrolled-jump-sw) rectangle (unrolled-jump-ne);
  \node[align=center] at ([xshift=2.1cm, yshift=0.65cm]unrolled.center) {Noisy samples around\\the unfolded manifold} ;
  \path[->, col2] let \p1 = (spiral-sw), \p2 = (spiral-ne), \p3 = (unrolled-sw), \p4 = (unrolled-ne) 
    in (\x1, \y1/2 + \y2/2) edge[out=210, in=90] node[above left=0, pos=0.55] {True neighbors} (\x3/2 + \x4/2, \y4);
  \path[->, col3] let \p1 = (spiral-jump-sw), \p2 = (spiral-jump-ne), \p3 = (unrolled-jump-sw), \p4 = (unrolled-jump-ne) 
    in (\x1/2 + \x2/2, \y1) edge[out=-90, in=90] node[below right=0, pos=0.75] {False neighbors} (\x3/2 + \x4/2, \y4);

  \begin{axis}[
    name=projected,
    yshift=-4.5cm,
    enlargelimits=false,
    ticks=none,
    axis lines=none,
    ymin=-0.8, ymax=0.8
  ]
  \addplot+[mark=none, color=black, smooth] table {\datapath/unrolled_spiral.data};
  \addplot+[col1, only marks, mark=*, mark size=1, opacity=0.5, mark options={fill=col1, color=col1}, y filter=/.code={0}] table[x=0, y=2] {\unrollednoisetable};
  \coordinate (projected-1) at (1.8876326887658077, 0);
  \coordinate (projected-2) at (1.9556583067767028, 0);
  \coordinate (projected-jump-1) at (2.12, 0);
  \coordinate (projected-jump-2) at (4.368817359432314, 0.0);
  \end{axis}
  
  \draw[col2, fill=col2] (projected-1) circle (1pt);
  \draw[col2, fill=col2] (projected-2) circle (1pt);
  \path let \p1 = (projected-1) in coordinate (projected-sw) at (\x1 - 2, \y1 - 2);
  \path let \p2 = (projected-2) in coordinate (projected-ne) at (\x2 + 2, \y2 + 2);
  \draw[col2] (projected-sw) rectangle (projected-ne);
  
  \draw[col3, fill=col3] (projected-jump-1) circle (1pt);
  \draw[col3, fill=col3] (projected-jump-2) circle (1pt);
  \path let \p1 = (projected-jump-1) in coordinate (projected-jump-sw) at (\x1 - 2, \y1 - 2);
  \path let \p2 = (projected-jump-2) in coordinate (projected-jump-ne) at (\x2 + 2, \y2 + 2);
  \draw[col3] (projected-jump-sw) rectangle (projected-jump-ne);

  \path[->, col2] let \p1 = (unrolled-sw), \p2 = (unrolled-ne), \p3 = (projected-sw), \p4 = (projected-ne) 
    in (\x1/2 + \x2/2, \y1) edge (\x3/2 + \x4/2, \y4);
  \path[->, col3] let \p1 = (unrolled-jump-sw), \p2 = (unrolled-jump-ne), \p3 = (projected-jump-sw), \p4 = (projected-jump-ne) 
    in (\x1/2 + \x2/2, \y1) edge (\x3/2 + \x4/2, \y4);
  \node[align=center] at ([xshift=2.1cm, yshift=0.3cm]projected.center) {1D embedding} ;

\end{tikzpicture}
    \caption{Example of dimensionality reduction. Top: Data samples 
    around an a priori unknown manifold.
    Middle: Illustrative "unfolded" manifold. 
    Bottom: An embedding of the dataset obtained by associating 1D coordinates to each sample.
    }
    \label{fig:spiral}
\end{figure}

Dimensionality reduction (DR) techniques are used in CC 
to 
(i) identify the CSI manifold within a CSI dataset and (ii) embed it into a low-dimensional chart.
Fig.~\ref{fig:spiral} illustrates a simple example in which the data lies on a spiral that is locally similar to a line, i.e., a 1D manifold embedded in 2D space.
DR strives to unfold the embedded manifold in the \emph{ambient} (here, 2D) space into its natural representation in the \emph{latent} (here, 1D) space.
DR has a rich history in 
statistical learning~\cite{VanDerMaaten2009}, where it is 
used to 
visualize datasets and to extract meaningful features from high-dimensional data.

DR algorithms may be classified, e.g., by 
\begin{itemize}
    \item whether they produce a parametric or non-parametric mapping 
    of samples in the ambient space to the latent space; 
    \item whether they estimate the intrinsic manifold dimension;
    \item whether they operate online on a stream of data, or offline on batches of data;
    \item whether they 
    distinguish between connected components in the underlying manifold;
    \item whether they output 
    layered embeddings, gradually  
    adding \emph{layers} or \emph{dimensions} until a desired metric is achieved, or not. 
\end{itemize}
One important issue which naturally arises in the dimensionality reduction process when applied to real-world datasets is the presence of noise (due to imperfect CSI estimation) in the input data samples, which can corrupt the manifold learning process.
As seen in Fig.~\ref{fig:spiral}, due to noise, the latent space is not truly 1D after unfolding. DR algorithms therefore 
project the data to 
limit the latent space dimension.
Noise may also introduce false neighbors, as exemplified in the figure, which 
may detrimentally effect
the success of the procedure if not handled properly;
many DR algorithms rely on creating local neighborhoods in the ambient space to learn the manifold.
Wrongly considering these two samples as neighbors would 
short circuit two folds of the spiral, and distort the inferred topology of the manifold.
Noise also impacts the estimation of the intrinsic dimensionality of the manifold, as can be intuitively understood from Fig.~\ref{fig:spiral}.
The noise variance may be larger than the variance of the manifold in 
some area, such that
the estimated dimensionality will  
reflect noise
dimensionality. 

As DR methods are 
self-supervised, 
it is difficult to define metrics for measuring  unfolding accuracy.
Most metrics rely on comparisons between the unfolded manifold and a reference space, with a known pairing between samples in the manifold and the reference spaces.
The reference space can be the ambient space, or 
the geographical location space if known by supervision.
Preservation of key topological or geometrical measures
between samples in the latent space and the reference space
is evaluated.
Common
topological metrics
are {\it trustworthiness} and {\it continuity}~\cite{VanDerMaaten2009}, which compare the neighborhoods of points between the two spaces.
Samples sharing the same neighbors in both spaces will score positively in both metrics; samples being neighbors in one space but not another will affect the 
metric negatively.
A meaningful geometrical metric is the Kruskal stress~\cite{Kruskal1964}, 
measuring how well distances between pairs of samples in the latent space match distances in the reference space.

\section{Wireless Channel Charting}
\label{sec:charting}

A generic scenario leveraging CC consists of
UEs moving in space and 
a BS 
implementing DR. The process is typically carried out in two phases.

In the first phase, the BS
learns the channel chart.
While the mobile UEs are transmitting pilot signals from different locations and time instants, the BS
uses the received pilot signals to estimate CSI samples.
These CSI estimates are 
converted into CSI features through a
feature extraction function; the resulting CSI features are stored in a database.
After 
constructing the CSI feature dataset, the BS
performs dimensionality reduction. 
For parametric
DR, one learns a channel charting function.
For nonparametric DR, one directly computes the coordinates of points in the channel chart for every CSI sample.
This process is self-supervised, i.e., it does not require any ground-truth location information from the mobile users.

In the second phase, the BS
uses the learned channel-charting function or the channel chart positions of the previous samples together with an out-of-sample extension method~\cite{AlTous2020} to map a new CSI sample to the coordinates of a point in the channel chart.
This new point is then made available in real time for 
any chart-based application.

The two-phase approach outlined above does not fully leverage the advantages of 
self-supervision. The continuous acquisition of CSI in a practical system would enable 
learning and updating the channel chart 
continuously. 
This would ensure that the channel chart is always up-to-date and accurately reflects long-term changes in the environment. 
The development of such life-long machine learning methods is 
an active research topic (see Sec.~\ref{sec:futuredirections}).

One of the most critical steps is the extraction of CSI features.
The ultimate goal of this step is to prepare the CSI samples such that dimensionality reduction is most effective, meaning that an accurate channel chart can be learned with practical dataset sizes and acceptable complexity.
First, CSI feature extraction ensures that only large-scale properties of the propagation environment are captured while small-scale fading and noise are ignored.
Effective CSI features may include angle-of-arrival, relative time-of-flight, or frequency selectivity/power-delay profiles.
The features should be sufficiently expressive to prevent aliasing caused by nonunique CSI features while unfolding the manifold.
Second, CSI feature extraction must mitigate the impact of hardware impairments 
at both Tx and Rx,
taking into account system and hardware components in order to ensure that the features are resilient to timing and synchronization errors, residual carrier frequency and sampling rate offset, phase noise, etc.
Methods bypassing feature extraction and directly processing CSI samples 
have been introduced recently;
however, they require significantly larger dataset sizes to deal with variations induced by hardware impairments.
Third, CSI feature extraction 
serves the purpose to reduce the CSI feature dimension with the goals of reducing dataset sizes and accelerating manifold learning.

\begin{figure}[tp]
    \centering

  \begin{NiceTabular}{ccc}
      & Map & Channel chart  \\
     \rotate Dataset 1 &  \includegraphics[width=0.5\columnwidth]{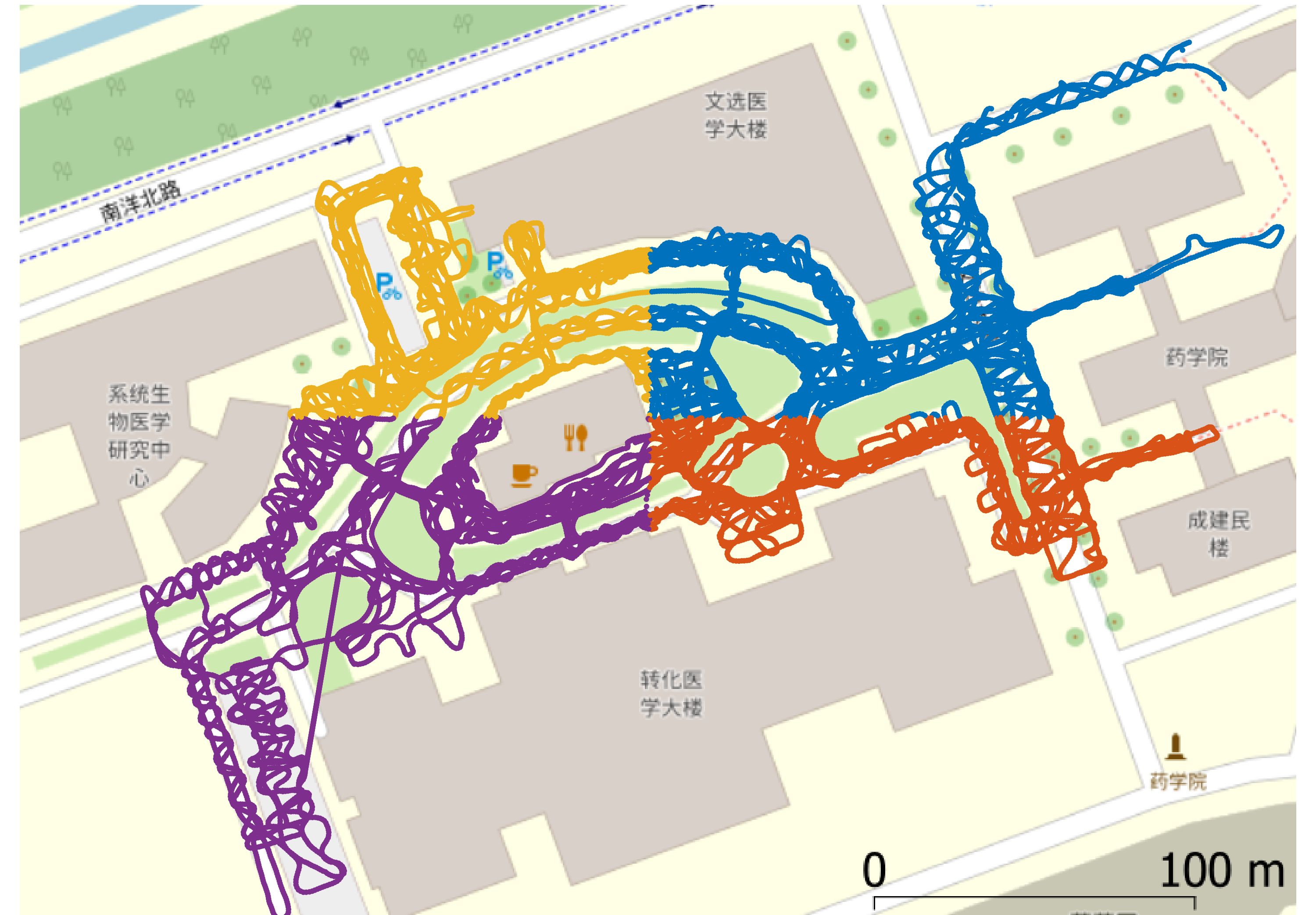} &  \includegraphics[width=0.35\columnwidth]{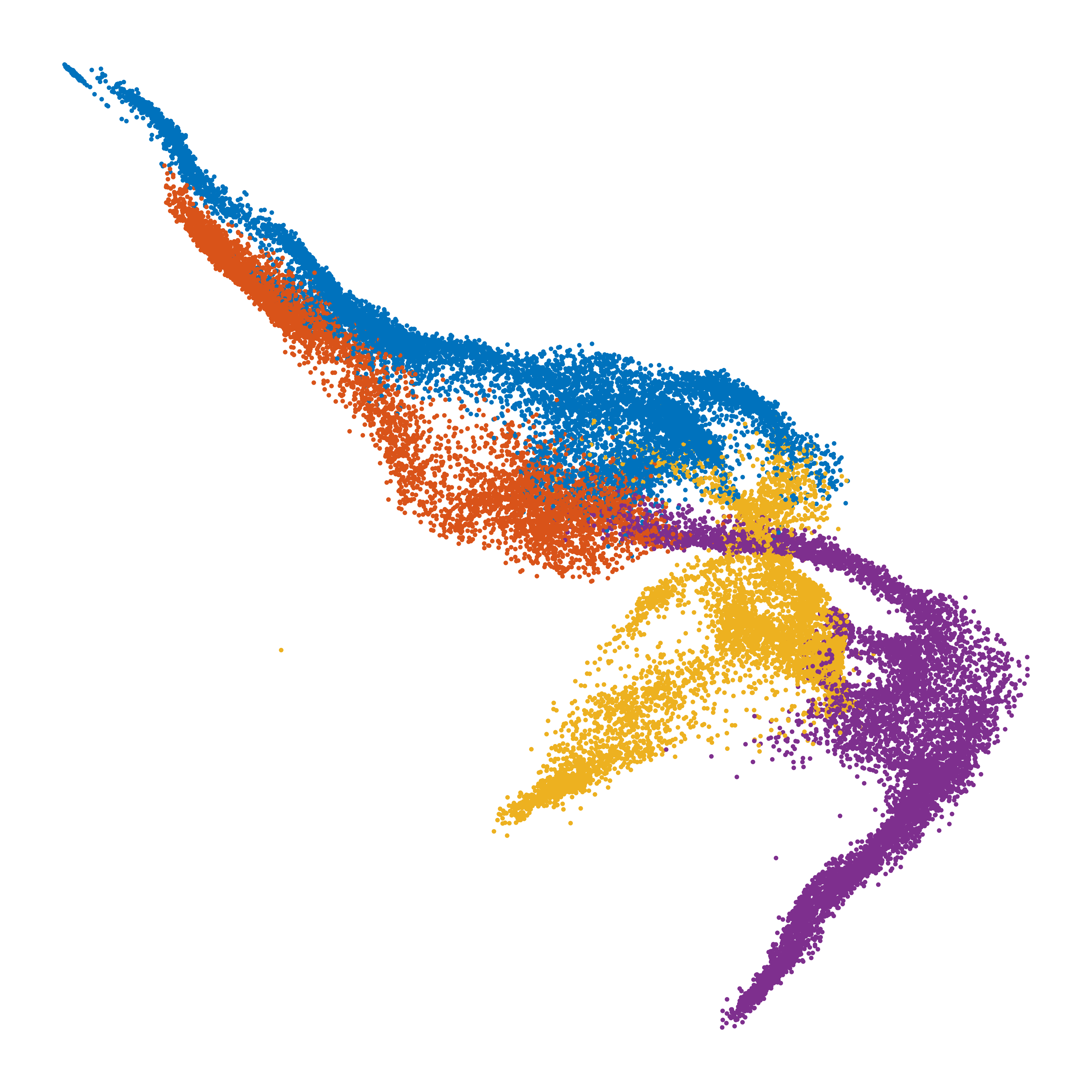}\\
     \rotate Dataset 2 &  \includegraphics[width=0.5\columnwidth]{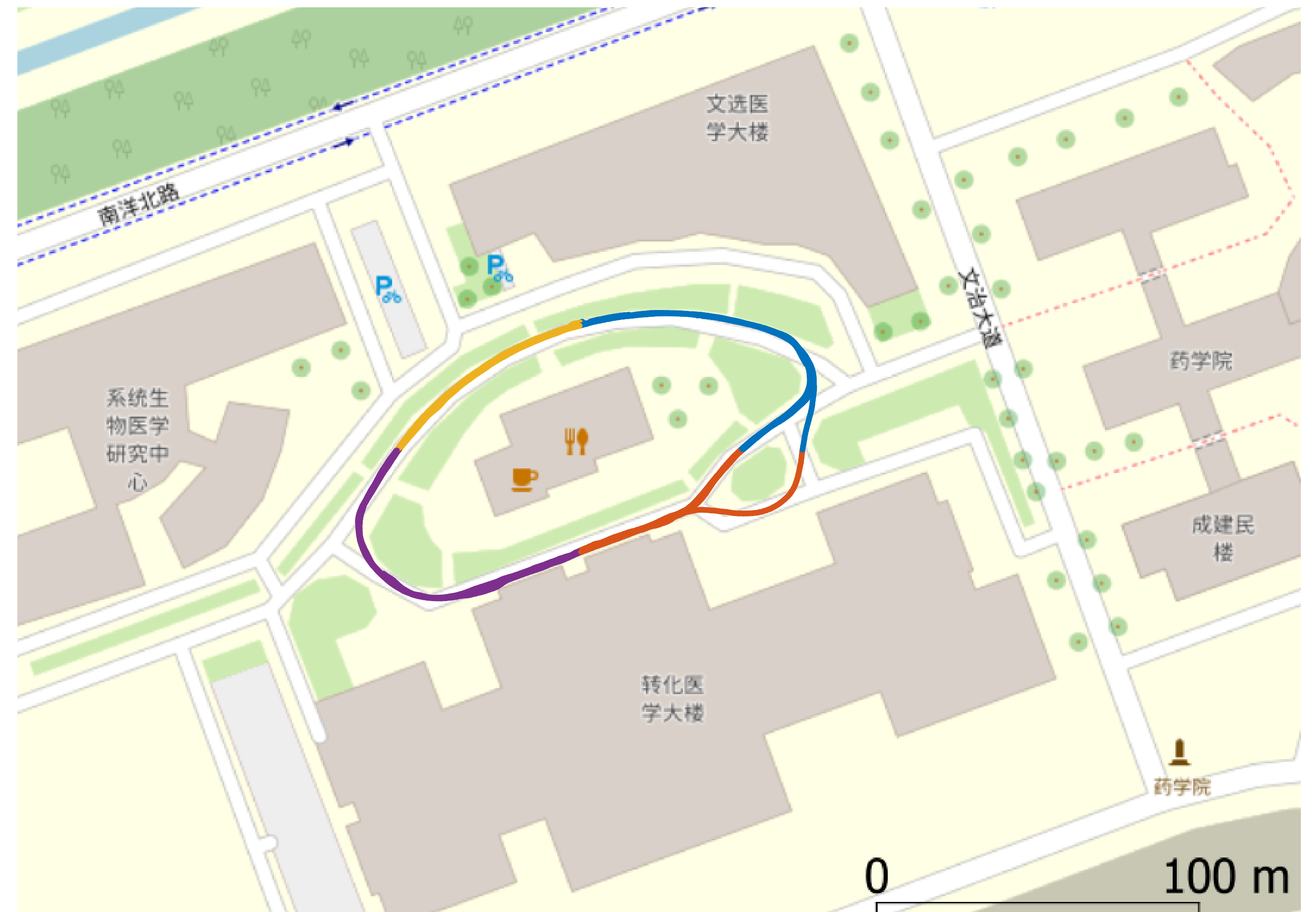} &  \includegraphics[width=0.35\columnwidth]{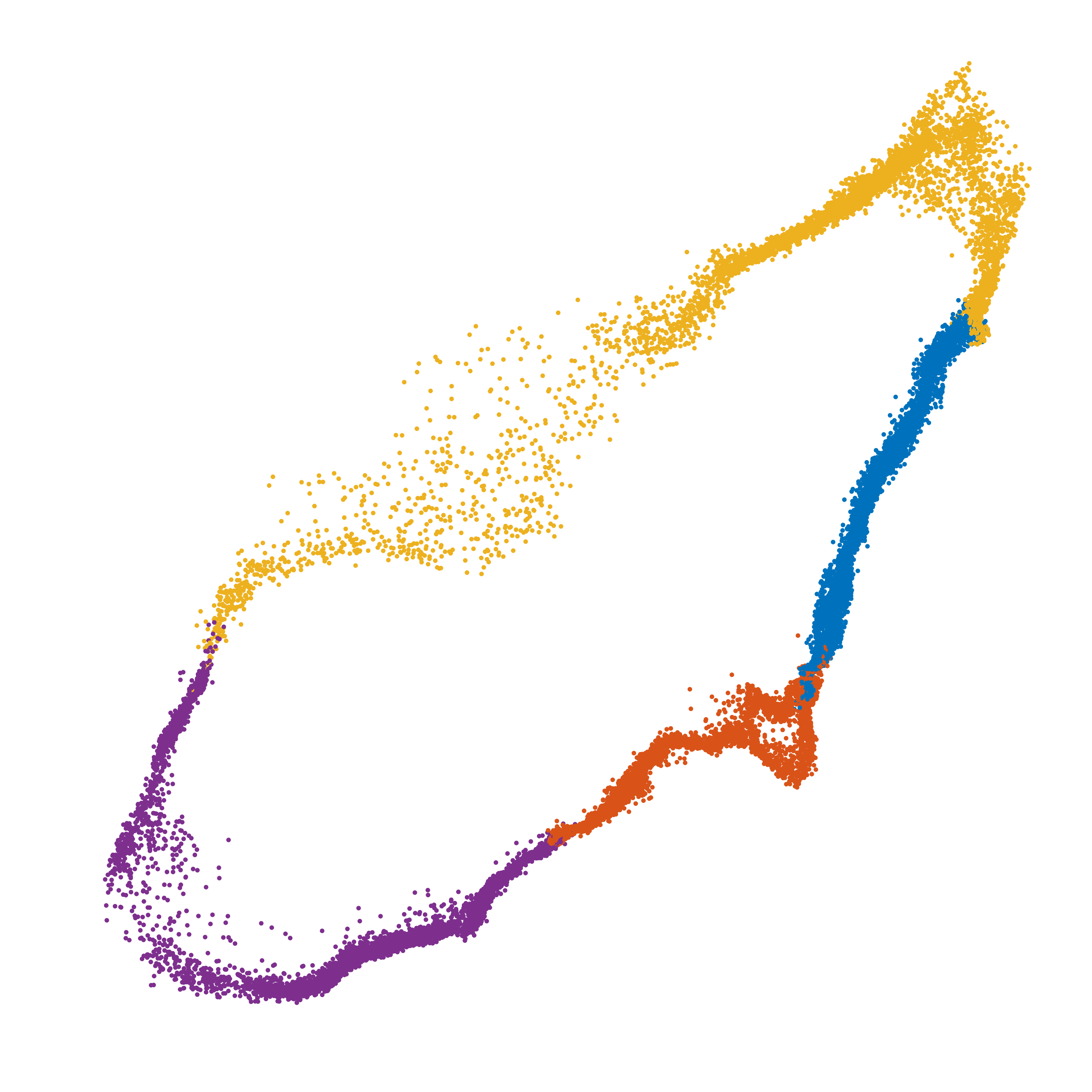}
   \end{NiceTabular}
   \caption{Channel charts from~\cite{Ferrand2021} for a mobile user moving in a campus area.
    Despite their self-supervised nature, the obtained channel charts (right) are topologically similar to the geographical positions (tracks with matching colors, left).}
    \label{fig:charting_results}
\end{figure}

\label{sec:channel_charting}

The other critical step in the first phase is dimensionality reduction. Many methods can be used for this for CC; 
parametric methods that rely on neural networks (NNs) have shown to be
particularly
effective~\cite{Ferrand2021}.
The reasons for their efficiency are as follows: First, 
NNs are powerful function approximators, which means that an explicit parametric mapping from CSI feature to a point in the channel chart can be learned, even for 
complicated channel manifolds.
They naturally allow for so-called \emph{out-of-sample} extension, whereby a new data point can be mapped to the low-dimensional latent space by means of the learned network.
Second, NNs
can be efficiently and accurately trained even for large datasets by leveraging stochastic gradient descent.
Powerful deep learning frameworks enable 
easy formulation of complex network architectures, loss functions, and regularizers, without having to implement DR  from scratch.
Third, one can readily utilize side information obtained during the acquisition process of CSI sampling to assist dimensionality reduction and improve the learned channel chart.
For example, in \cite{Ferrand2021} 
sample timestamp information 
is used to improve channel charting.
Generally, any proximity information can be used to regularize 
NN training and significantly improve the accuracy of the learned channel chart~\cite{Studer2018}.

In order for channel charting to be maximally effective, CSI feature extraction and  dimensionality reduction
should be   \emph{jointly} designed, and adapted to the communication system and Tx/Rx hardware.
%

\section{Experimental Results}
\label{sec:experiments}

Channel charting has been successfully applied to both synthetic channels~\cite{Studer2018,Agostini2020,AlTous2020,Kazemi2021,Deng2018} and real-world CSI measurements~\cite{Ferrand2021}, using 
DR algorithms, such as Sammon's mapping or Laplacian eigenmaps 
(see \cite{VanDerMaaten2009}),
as well as 
NN approaches, such as Siamese or triplet networks.

Fig.~\ref{fig:charting_results} depicts two channel charts from~\cite{Ferrand2021} obtained by training a triplet-based NN
on real-world CSI measurements.
The datasets were acquired using a commercial 4G base station, with a rooftop antenna array with 32 dual-polarized antenna elements in a uniform $8\times 4$ rectangular array.
The UE has a single antenna.
Orthogonal frequency division multiplexing with 288 subcarriers in a 10\,MHz bandwidth at 2.5\,GHz is used.
Dataset 1 includes
some 3 hours of pedestrian movement---at about 1 m/s---around the dense urban coverage area, with both line-of-sight (LoS) and non-LoS areas, amounting to several kilometers worth of tracks.
The resulting channel chart 
shows excellent connectivity of the four colored subsets,
but also 
illustrates the inherent uncertainty about rotations and mirroring within a channel chart when compared to the geographical position.
Dataset 2 consists in a 20-minute run where the UE was moved multiple times along an approximately elliptic trajectory around the square.
The resulting channel chart correctly ``connects'' the loop as the meaningful representation of the CSI dataset.
These results show that channel charting is able to recover 
topological properties from measured data in a completely self-supervised manner.
Both channel charts in Fig.~\ref{fig:charting_results} score well with respect to the metrics discussed in Sec.~\ref{sec:dimreduction}. A complete numerical performance assessment is available in~\cite{Ferrand2021}.

\begin{table*}[h]
\centering
\caption{Applications of Channel Charting}
\label{table_applications}
\resizebox{2\columnwidth}{!}{%
\begin{tabular}{p{4cm}p{8cm}p{6cm}@{}}
  \toprule
  \bf Application & \bf Details & \bf Leveraged channel chart property \\
  \midrule
  Radio resource management & Handover prediction~\cite{channelchartingresources_website}, pilot allocation~\cite{Ribeiro2020}, rate or protocol adaptation, predictive buffering~\cite{Kasparick_TVT2016}, BS association. & Temporal consistency, user trajectory extrapolation\\
  \midrule
  mmWave beam management & Accelerated beam discovery~\cite{Kazemi2021}, predictive beam tracking  &  Temporal and spatial consistency, multi-view chart\\
  \midrule
  Device grouping for D2D & Identify and cluster nearby mobile devices~\cite{AlTous2020} & Spatial consistency\\
  \midrule
  CSI compression & Representing CSI using the minimal number of dimensions, for efficient quantization and feedback  & Data-driven, time consistency\\
  \midrule
  Proximity detection, geofencing & Detect user proximity or relative position to a fixed feature on the ground (shop, prohibited areas) using the channel chart & Spatial consistency\\
  \midrule
  Network event labeling & Identify and cluster link failures events and low performance spots for network planning purposes & Spatial and temporal  consistency\\
  \midrule
  Context awareness & Infer contextual information (e.g., pedestrian or in-vehicle; work commute or unusual trip) from the radio link & Data driven, spatial and temporal consistency\\
  \midrule
  Localization & Supervised or semi-supervised version of charting \cite{channelchartingresources_website} & Spatial and temporal consistency\\
  \midrule
  Channel modeling and simulation & Understand and model the wireless propagation medium, generate realistic CSI samples & Data-driven, temporal consistency\\
 \bottomrule
\end{tabular}}
\end{table*}

\section{Applications}
\label{sec:applications}

A channel chart, being  merely  
a low-dimensional representation of the state of the wireless channel, 
is not a goal in itself. 
Instead, it should be seen as a versatile platform upon which 
applications can rely to infer, predict, and process propagation-related information. \revision{We now showcase two example applications that demonstrate the efficacy of CC.}

\begin{figure}[tp]
	\centering
	\includegraphics[width=0.75\columnwidth]{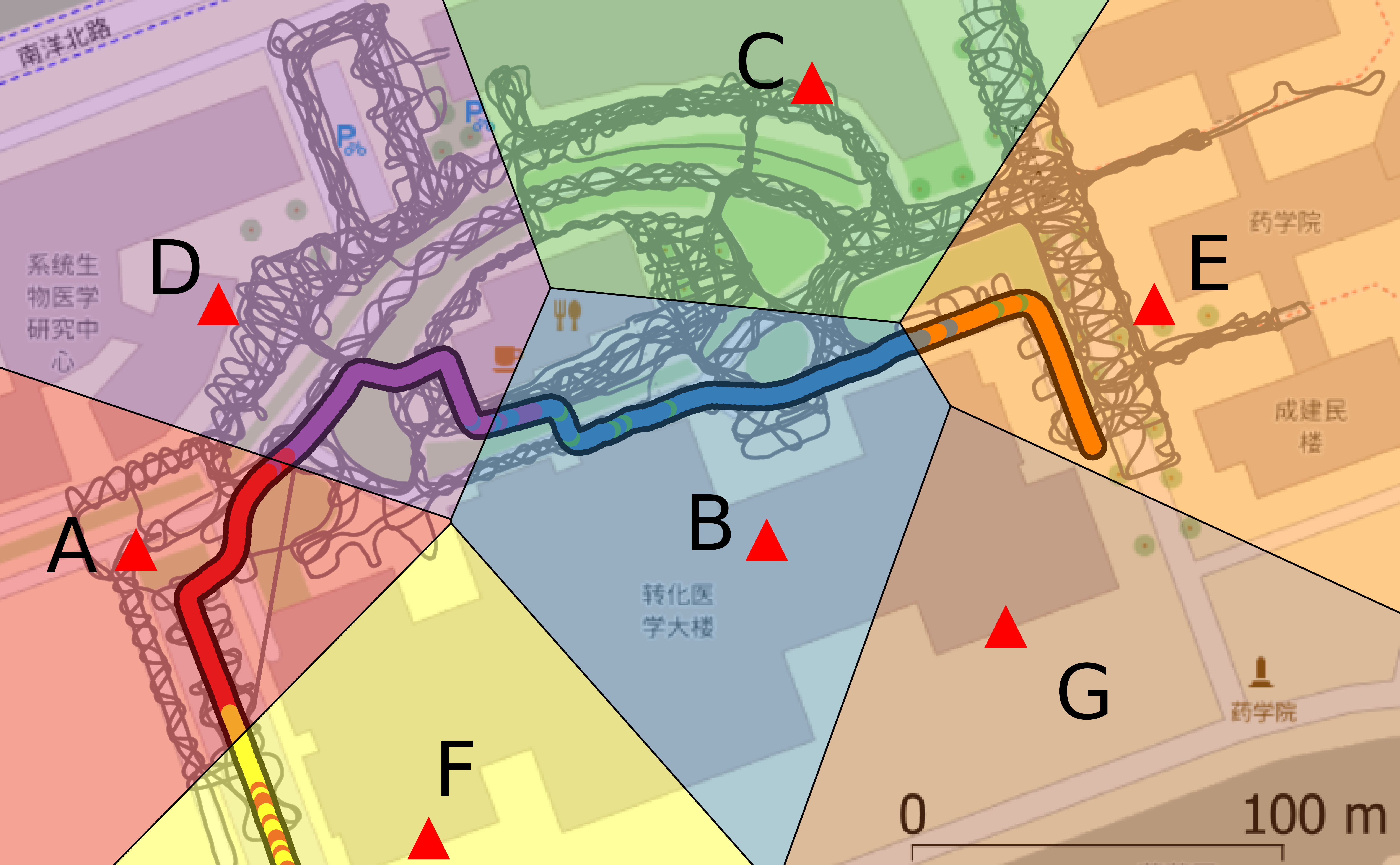}
	 \centering
	 \includegraphics[width=0.8\columnwidth]{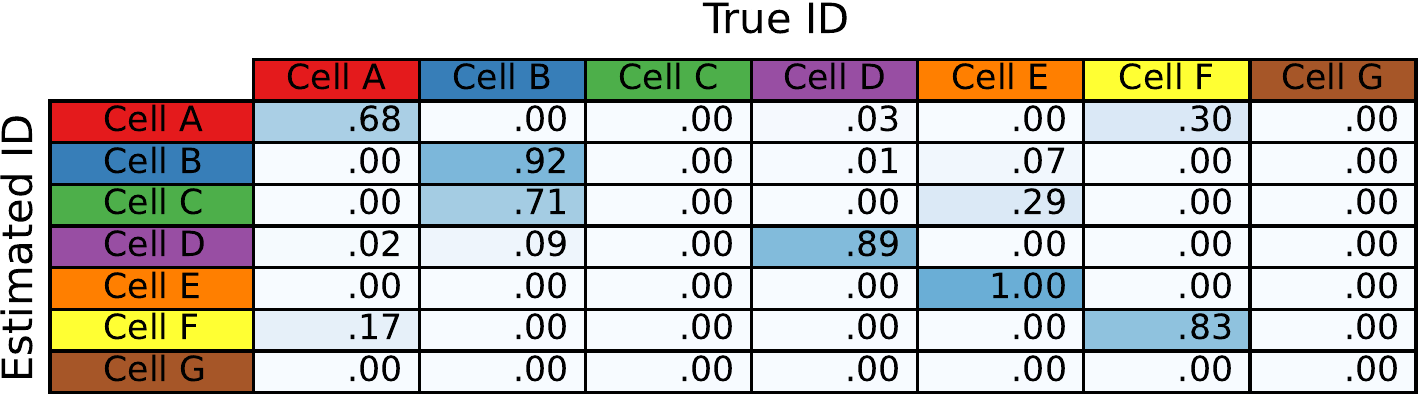}
    
    \caption{Example of chart-based cell association in a network of seven mmWave cell. (Top) test user trajectory colored according to the estimated cell ID. The channel chart (not shown) was constructed from Dataset 1, shown in gray. (Bottom) confusion matrix showing the probability of the true cell ID, for each estimated cell ID.}
    \label{fig:chart-classification}
\end{figure}

\revision{The first example deals with mmWave beam discovery in a dual-band set-up.}
A channel chart is constructed for Dataset 1 from Fig.~\ref{fig:charting_results}, for the massive MIMO CSI.
The associated cell connectivity for the mmWave band is simulated as depicted in Fig.~\ref{fig:chart-classification}. CSI samples in the channel chart corresponding to the training set are
colored according to their proximity with mmWave BSs (red triangles).
Samples from the test dataset (trajectory depicted in the figure) are mapped to the channel chart, and their mmWave cell association is estimated
from the color of the neighboring training set points in the channel chart.
\revision{This approach achieves 85\% accuracy;  most errors occur at cell-edge locations where a neighboring cell is chosen instead.} 

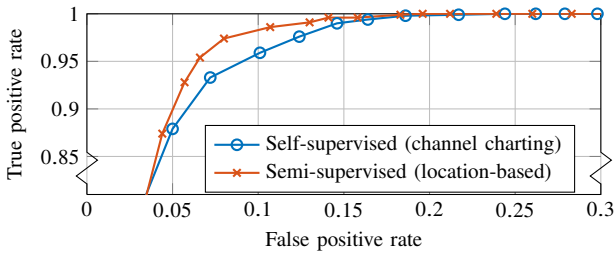
\begin{figure}[htp]
     \centering
      \footnotesize
      \begin{tikzpicture}
        \begin{axis}[
          width=0.95\columnwidth,
          height=0.45\columnwidth,
          xlabel={False positive rate},
          ylabel={True positive rate},
          xmin=0, xmax=0.3,
          ymin=0.81, ymax=1,
          xtick={0, 0.05, 0.1, 0.15, 0.2, 0.25, 0.3},
          tick label style={/pgf/number format/fixed},
          axis y discontinuity=crunch,
          grid=both,
          legend style={at={(0.95,0.01)}, anchor=south east},
          legend cell align={left},
        ]
    
          \addplot[thick, color=col1, thick, mark=o] table[x=fpr, y=tpr, col sep=comma] {figures/jsac_roc.csv};
          \addlegendentry{Self-supervised (channel charting)}
          \addplot[thick, color=col2, thick, mark=x] table[x=fpr, y=tpr, col sep=comma] {figures/jsac_roc_anchors.csv};
          \addlegendentry{Semi-supervised (location-based)}
        \end{axis}
      \end{tikzpicture}
      \caption{\revision{ROC curves for channel-chart-based proximity detection.}}
      \label{fig:proximity_roc}
\end{figure} 
\revision{The second example 
deals with proximity detection. 
We train two channel charts using Dataset~1 again: a pseudo-location based fully self-supervised channel chart and a semi-supervised version using known Tx-position information acquired every five minutes.
Test user positions are sampled from the channel chart, 
and their position is averaged over 
five seconds.
Pairs of test users are declared neighbors if their distance in the channel chart is lower than some threshold and non-neighbors otherwise.
The receiver operating characteristic (ROC) curves of this hypothesis test curves depicted in Fig.~\ref{fig:proximity_roc} indicate that it is 
possible to achieve a high true positive detection rate with few false positives.
For example, a 10\% false positive rate yields a true positive rate of 96\% for the self-supervised and 98\% for the semi-supervised case.
Pseudo-positioning via CC can thus replace positioning with minimal accuracy loss in this proximity detection example.}

\revision{Beyond the two applications 
above, Table \ref{table_applications} lists a 
range of potential applications of CC and identifies the relevant channel chart properties.
}
Several of the applications in this table can be solved without 
resorting to a channel chart. For instance, the correspondence between mmWave beam direction and the CSI measured on a low-frequency band can be learned directly.
However, a channel chart provides a simple 
unified representation of propagation-related characteristics which may be shared among several applications using a standardized interface,
thus allowing significant savings in computation and storage over parallel end-to-end solutions each involving a separate learning problem. Furthermore, the CC-based solution does not require 
directly exposing CSI or 
proprietary base-station hardware details.
In this ``\emph{channel charting as a service}'' paradigm, the computationally-intensive channel chart update and maintenance tasks are delegated to the host network, while simple but powerful access to propagation-related characteristics can be extended up to the application layer, enabling the ex-post development of innovative applications not listed in Table~\ref{table_applications}.

\section{Challenges, Extensions,  and Future Directions}
\label{sec:futuredirections}

A successful implementation of channel charting presents unique challenges beyond solving the DR problem.
\subsubsection{Continuous learning}
The channel chart 
should evolve over time, reflecting changes in the propagation environment. Some 
changes 
may be permanent (e.g., construction of a new building) and will require a channel chart update, while others might be temporary, for instance the presence a large vehicle altering propagation. In the former case, past information shall be forgotten, while in the latter case, the presence/absence of the vehicle might be captured as extra dimension of the channel chart.
It would then constitute information that could be inferred from CSI and used for context-aware applications. 
Alternatively, the channel chart can be made robust to such changes, considering them a nuisance rather than useful information.

\subsubsection{Online and real-time aspects}
CSI typically streams in continuously at a rate of 
hundreds samples per second.
This situation departs from 
classical machine learning 
problems where fixed-size datasets are processed offline. 
Algorithmic adaptations are required to achieve online, life-long learning.
In CC applications, both learning (charting) and inference needs to be performed with stringent real-time and storage constraints, which opens up numerous challenges related to the real-time implementation of machine learning algorithms for the physical layer.

\subsubsection{Distributed implementation}
In modern radio access networks,  mobile devices can 
be served by multiple physically distinct transmission points (TRPs) and in multiple frequency bands.
In this context, the capability of  constructing joint or distributed channel charts across multiple TRPs and bands, i.e., \emph{multi-view} channel charts~\cite{Deng2018} will be a key enabler for handover and beam steering applications.

\subsubsection{Array geometry and propagation scenario}
CC for indoor applications presents unique challenges: the keyhole effect typically observed in outdoor-to-indoor propagation scenarios might limit the expressivity of the channel chart, while the relevance of charting for indoors systems will hinge on a favorable geometry of the antenna system (e.g., data gathered from a distributed antenna system will capture a richer range of environmental properties than a single array with co-located antennas).
Analog Tx- and/or Rx-side beamformers are commonly used during CSI estimation, and they are required in high frequency bands in order to guarantee a sufficient link budget---this yields a partial channel observation that complicates channel charting.
A proper framework for handling time-varying beamformers within the charting toolchain is necessary~\cite{Kazemi2021}.
In this context, closed-loop strategies (such as reinforcement learning approaches) involving dynamic adjustments of the CSI measurement process can potentially further enhance the chart coverage and accuracy.

\subsubsection{Privacy aspects}
Some of the applications described in Table~\ref{table_applications} (e.g., proximity detection) reveal that the channel chart's pseudo-position information may have to be
considered under the angle of privacy protection.
Note that this is generally true for RANs which have provided some kind of localization services since 2G.
With the increasing importance of proper handling of private information, it is expected that 6G will have built-in technical features guaranteeing privacy through a powerful data governance architecture for better societal acceptance \cite[Chap.~30]{tong_zhu_2021}.

\bibliographystyle{IEEEtran}
\bibliography{IEEEabrv,charting}

\begin{thebibliography}{10}
\providecommand{\url}[1]{#1}
\csname url@samestyle\endcsname
\providecommand{\newblock}{\relax}
\providecommand{\bibinfo}[2]{#2}
\providecommand{\BIBentrySTDinterwordspacing}{\spaceskip=0pt\relax}
\providecommand{\BIBentryALTinterwordstretchfactor}{4}
\providecommand{\BIBentryALTinterwordspacing}{\spaceskip=\fontdimen2\font plus
\BIBentryALTinterwordstretchfactor\fontdimen3\font minus
  \fontdimen4\font\relax}
\providecommand{\BIBforeignlanguage}[2]{{%
\expandafter\ifx\csname l@#1\endcsname\relax
\typeout{** WARNING: IEEEtran.bst: No hyphenation pattern has been}%
\typeout{** loaded for the language `#1'. Using the pattern for}%
\typeout{** the default language instead.}%
\else
\language=\csname l@#1\endcsname
\fi
#2}}
\providecommand{\BIBdecl}{\relax}
\BIBdecl

\bibitem{Kasparick_TVT2016}
M.~Kasparick, R.~L.~G. Cavalcante, S.~Valentin, S.~Sta\'nczak, and M.~Yukawa,
  ``Kernel-based adaptive online reconstruction of coverage maps with side
  information,'' \emph{IEEE Trans. Veh.}, vol.~65, no.~7, pp. 5461--5473, Jul.
  2016.

\bibitem{tong_zhu_2021}
W.~Tong and P.~Zhu, Eds., \emph{6G: The Next Horizon: From Connected People and
  Things to Connected Intelligence}.\hskip 1em plus 0.5em minus 0.4em\relax
  Cambridge University Press, 2021.

\bibitem{Alamu2021}
O.~Alamu, B.~Iyaomolere, and A.~Abdulrahman, ``{An overview of massive MIMO
  localization techniques in wireless cellular networks: Recent advances and
  outlook},'' \emph{Elsevier Ad Hoc Netw.}, vol. 111, Feb. 2021.

\bibitem{Studer2018}
C.~Studer, S.~Medjkouh, E.~Gönültaş, T.~Goldstein, and O.~Tirkkonen,
  ``Channel charting: Locating users within the radio environment using channel
  state information,'' \emph{IEEE Access}, vol.~6, pp. 47\,682--47\,698, Aug.
  2018.

\bibitem{VanDerMaaten2009}
L.~{Van Der Maaten}, E.~Postma, and J.~{Van den Herik}, ``Dimensionality
  reduction: a comparative review,'' \emph{J. Mach. Learn. Res.}, vol.~13,
  no.~10, pp. 66--71, Oct. 2009.

\bibitem{Bourdoux2020}
{A. Bourdoux \& al.}, ``{6G} white paper on localization and sensing,''
  University of Oulu, Tech. Rep., Jun. 2020, {ISBN} 978-952-62-2674-3.

\bibitem{Kazemi2021}
P.~Kazemi, T.~Ponnada, H.~Al-Tous, Y.-C. Liang, and O.~Tirkkonen, ``Channel
  charting based beam {SNR} prediction,'' in \emph{Proc. Eur. Conf. on Netw.
  Com.}, Porto, Portugal, Jun. 2021.

\bibitem{Ribeiro2020}
L.~Ribeiro, M.~Leinonen, D.~Djelouat, and M.~Juntti, ``Channel charting for
  pilot reuse in {mMTC} with spatially correlated {MIMO} channels,'' in
  \emph{Intl. Conf. Cognitive Radio Oriented Wireless Netw.}, Taipei, Taiwan,
  Dec. 2020.

\bibitem{Ferrand2021}
P.~Ferrand, A.~Decurninge, L.~G. Ordo{\~n}ez, and M.~Guillaud, ``Triplet-based
  wireless channel charting: Architecture and experiments,'' \emph{{IEEE} J.
  Sel. Areas Commun.}, vol.~39, no.~8, pp. 2361--2373, Aug. 2021.

\bibitem{channelchartingresources_website}
``Channel charting resources,'' website
  \url{https://channelcharting.github.io/}.

\bibitem{Schmidt1986}
R.~Schmidt, ``Multiple emitter location and signal parameter estimation,''
  \emph{IEEE Trans. Ant. and Propag.}, vol.~34, no.~3, pp. 276--280, Mar. 1986.

\bibitem{Kruskal1964}
J.~B. Kruskal, ``Multidimensional scaling by optimizing goodness of fit to a
  nonmetric hypothesis,'' \emph{Psychometrika}, vol.~29, no.~1, Mar. 1964.

\bibitem{AlTous2020}
H.~Al-Tous, T.~Ponnada, C.~Studer, and O.~Tirkkonen, ``Multipoint channel
  charting-based radio resource management for {V2V} communications,''
  \emph{EURASIP J. Wireless Commun. Netw.}, Dec. 2020.

\bibitem{Agostini2020}
P.~Agostini, Z.~Utkovski, and S.~Stańczak, ``Channel charting: An {Euclidean}
  distance matrix completion perspective,'' in \emph{IEEE Intl. Conf. Acoust.,
  Speech, Signal Process.}, Barcelona, Spain, May 2020.

\bibitem{Deng2018}
J.~Deng, S.~Medjkouh, N.~Malm, O.~Tirkkonen, and C.~Studer, ``Multipoint
  channel charting for wireless networks,'' in \emph{Proc. Asilomar Conf.
  Signals, Syst., Comput.}, Pacific Grove, CA, USA, Oct. 2018.

\end{thebibliography}

\begin{IEEEbiography}{Paul Ferrand}
received the M.Sc. degree in computer science from the Université de Lyon, France, in 2009, and the Ph.D. degree from INSA Lyon, France, in 2013. He was a Post-Doctoral Fellow with INRIA, France, where he was involved in energy-efficient communications.
From 2014 to 2021, he was a Senior Researcher at the Huawei Research Center, Paris, France. His main research interests lie within the field of signal and information processing for the physical layer of wireless communications.
\end{IEEEbiography}

\begin{IEEEbiography}{Maxime Guillaud}
is a senior researcher at Inria in Lyon.
He has over 20 years of expertise in the domain of wireless communications in both academic and industrial research environments.
He received his Ph.D. degree in electrical engineering and communications from EURECOM and Telecom ParisTech in 2005. From 2006 to 2010, he was a Senior Researcher with FTW. From 2010 to 2014, he was a research associate at Vienna University of Technology. From 2014 to 2022, he was with Huawei Technologies. His expertise lies the physical layer of RANs, including transceiver algorithms, channel modeling, and modulation design.
\end{IEEEbiography}

\begin{IEEEbiography}{Christoph Studer}
is an Associate Professor at the Department of Information Technology and Electrical Engineering at ETH Zurich in Switzerland.
Dr.\ Studer received his M.S.\ and Ph.D.\ degrees in electrical engineering from ETH Zurich in 2005 and 2009, respectively. Between 2009 and 2013, he held postdoctoral positions at ETH Zurich and Rice University in Houston, TX.
In 2014, he joined Cornell University, Ithaca, NY, as an Assistant Professor. From 2019 to 2020, he was an Associate Professor at Cornell University and Cornell Tech in New York City. In 2020, he joined ETH Zurich. 
Dr.\ Studer's research interests include the design of digital integrated circuits, as well as wireless communications, digital signal processing,  numerical optimization, and machine learning.
\end{IEEEbiography}

\begin{IEEEbiography}{Olav Tirkkonen}
 is a professor in communication theory at Aalto University, Finland, where he has held a faculty position since  2006. He received his M.Sc. and Ph.D. degrees in theoretical physics from Helsinki University of Technology in 1990 and 1994, respectively. He held post-doctoral positions at UBC, Vancouver, Canada, and NORDITA, Copenhagen, Denmark, and was with Nokia Research Center, Helsinki, Finland from 1999 to 2010. In 2016--2017, he was Visiting Associate Professor at Cornell University, Ithaca, NY, USA.
His current research interests are in coding for random access and quantization, quantum computation, and machine learning for cellular networks.
\end{IEEEbiography}

\end{document}